\documentclass[]{aa}

\usepackage{txfonts}

\usepackage[english]{babel}
\usepackage{amsmath,amssymb}
\usepackage{graphicx, subfig}
\usepackage[normalem]{ulem}

\usepackage{verbatim}

\usepackage{multirow}
\usepackage{mathtools}
\usepackage{epstopdf}

\usepackage{url}
\usepackage{xcolor,color}
\usepackage{orcidlink}

\usepackage{footmisc}
\usepackage{natbib,twoopt}
\usepackage[hyphenbreaks]{breakurl}

\bibpunct{(}{)}{;}{a}{}{,}             
\definecolor{cobalt}{rgb}{0.06, 0.2, 0.65}
\hypersetup{
  colorlinks,
  citecolor=cobalt,
  linkcolor=[rgb]{0.8, 0.2, 1.0},
  urlcolor=cobalt,
}
\makeatletter
  \newcommandtwoopt{\citeads}[3][][]{\href{http://adsabs.harvard.edu/abs/#3}%
    {\def\hyper@linkstart##1##2{}%
     \let\hyper@linkend\@empty\citealp[#1][#2]{#3}}}
  \newcommandtwoopt{\citepads}[3][][]{\href{http://adsabs.harvard.edu/abs/#3}%
    {\def\hyper@linkstart##1##2{}%
     \let\hyper@linkend\@empty\citep[#1][#2]{#3}}}
  \newcommandtwoopt{\citetads}[3][][]{\href{http://adsabs.harvard.edu/abs/#3}%
    {\def\hyper@linkstart##1##2{}%
     \let\hyper@linkend\@empty\citet[#1][#2]{#3}}}
  \newcommandtwoopt{\citeyearads}[3][][]%
    {\href{http://adsabs.harvard.edu/abs/#3}
    {\def\hyper@linkstart##1##2{}%
     \let\hyper@linkend\@empty\citeyear[#1][#2]{#3}}}
\makeatother

\newcommand\code[1]{\textsc{\MakeLowercase{#1}}}
\newcommand{\quotes}[1]{``#1''}

\def\msun{{\rm M}_{\odot}}
\def\zsun{{\rm Z}_{\odot}}

\def\pc{{\rm {pc}}}

\def\angstrom{\textrm{A\kern -1.3ex\raisebox{0.6ex}{$^\circ$}}}

\def\myr{{\rm Myr}}

\def\msunyr{\msun\,{\rm yr}^{-1}}

\def\be{\begin{equation}} 
\def\ee{\end{equation}} 
\def\ba{\begin{eqnarray}} 
\def\ea{\end{eqnarray}} 
\def\gtsima{$\; \buildrel > \over \sim \;$}
\def\ltsima{$\; \buildrel < \over \sim \;$}
\def\gsim{\lower.5ex\hbox{\gtsima}} 
\def\lsim{\lower.5ex\hbox{\ltsima}}
\def\prosima{$\; \buildrel \propto \over \sim \;$} 
\def\simgt{\lower.5ex\hbox{\gtsima}} 
\def\simlt{\lower.5ex\hbox{\ltsima}} 
\def\simpr{\lower.5ex\hbox{\prosima}}

\definecolor{mkcolor}{HTML}{01abdf}
\definecolor{apcolor}{HTML}{b3003b}
\definecolor{afcolor}{HTML}{01bdff}
\definecolor{lvcolor}{HTML}{ff9933}
\definecolor{sgcolor}{HTML}{9933ff}

\def\apjl{ApJL}
\def\apjs{ApJS}
\def\aap{A\&A}

\graphicspath{{plots_png/},{plots/}}

\def\tdelay{t_{\rm d}}
\def\acosmo{A_{100}}
\def\sigmasfr{\sigma_{\rm SFR}}
\def\sigmaz{\sigma_{Z}}

\begin{document}

\title{The mass-metallicity relation as a ruler for galaxy evolution:\\ insights from the James Webb Space Telescope}
\titlerunning{The MZR as a ruler for galaxy evolution: insights from the JWST}

\author{
A. Pallottini \orcidlink{0000-0002-7129-5761} \inst{1,2}\fnmsep\thanks{\href{mailto:andrea.pallottini@unipi.it}{andrea.pallottini@unipi.it}}
\and A. Ferrara \orcidlink{0000-0002-9400-7312} \inst{1}
\and S. Gallerani \orcidlink{0000-0002-7200-8293} \inst{1}
\and L. Sommovigo \orcidlink{0000-0002-2906-2200} \inst{3}
\and S. Carniani \orcidlink{0000-0002-6719-380X} \inst{1}
\and L. Vallini \orcidlink{0000-0002-3258-3672} \inst{4}
\and M. Kohandel \orcidlink{0000-0003-1041-7865} \inst{1}
\and G. Venturi \orcidlink{0000-0001-8349-3055} \inst{1}
}
\authorrunning{Pallottini et al.}

\institute{
Scuola Normale Superiore, Piazza dei Cavalieri 7, 56126 Pisa, Italy
\and Dipartimento di Fisica ``Enrico Fermi'', Universit\'{a} di Pisa, Largo Bruno Pontecorvo 3, Pisa I-56127, Italy
\and Center for Computational Astrophysics, Flatiron Institute, 162 5th Avenue, New York, NY 10010, USA
\and INAF-Osservatorio di Astrofisica e Scienza dello Spazio, via Gobetti 93/3, I-40129, Bologna, Italy
}
\date{Received July 31, 2024; accepted May 9, 2025}

\abstract
{
Galaxy evolution emerges from the balance between cosmic gas accretion, fueling star formation, and supernova feedback regulating metal enrichment of the interstellar medium. Hence, the relation between stellar mass ($M_\star$) and gas metallicity ($Z_g$) is fundamental to understanding the physics of galaxies. High-quality spectroscopic JWST data enable accurate measurements of both $M_\star$ and $Z_g$ up to redshift $z\simeq 10$.
}
{
Our aims are to understand \textit{(i)} the nature of the observed mass-metallicity relation (MZR), \textit{(ii)} its connection with the star formation rate (SFR), \textit{(iii)} the role played by SFR stochasticity (flickering), and \textit{(iv)} how it is regulated by stellar feedback.
}
{
We compare the MZR obtained by the JADES, CEERS, and UNCOVER surveys, which comprise about 180 galaxies at $z\simeq 3-10$ with $10^6 \msun \lsim M_\star \lsim 10^{10} \msun$, with $\simeq 200$ simulated galaxies in the same mass range from the \code{serra} high-resolution ($\simeq 20\,\rm \pc$) suite of cosmological radiation-hydrodynamic simulations.
To interpret the MZR, we develop a minimal, physically motivated, model for galaxy evolution that includes: cosmic accretion, possibly modulated with an amplitude $\acosmo$ on $100\,\myr$ timescales; a time delay $\tdelay$ between SFR and supernova feedback; SN-driven outflows with a varying mass loading factor $\epsilon_{\rm SN}$, that is normalized to the \code{FIRE} simulations predictions for $\epsilon_{\rm SN}=1$.
}
{
Using our minimal model, we find the observed \textit{mean} MZR is reproduced for relatively inefficient outflows ($\epsilon_{\rm SN}=1/4$), in line with findings from JADES.
Matching the observed MZR \textit{dispersion} across the full stellar mass range requires a delay time $\tdelay=20\,\myr$, in addition to a significant modulation ($\acosmo = 1/3$) of the accretion rate.
Successful models are characterized by a relatively low flickering ($\sigmasfr\simeq 0.2$), corresponding to a metallicity dispersion of $\sigmaz\simeq 0.2$. Such values are close but slightly lower than predicted from \code{serra} ($\sigmasfr\simeq 0.24$, $\sigmaz\simeq 0.3$), clarifying why \code{serra} shows a flatter trend with respect to the observations and some tension, especially at $M_\star \simeq 10^{10} \msun$.
}
{
The MZR appears to be very sensitive to SFR stochasticity, the minimal model predicts that high r.m.s. values ($\sigmasfr \simeq 0.5$) result in a \quotes{chemical chaos} (i.e. $\sigmaz\simeq 1.4$), virtually destroying the observed MZR. As a consequence, invoking a highly stochastic SFR ($\sigmasfr \simeq 0.8$) to explain the overabundance of bright, super-early galaxies would lead to inconsistencies with the observed MZR. 
}

\maketitle

\section{Introduction}\label{sec:intro}

The baryon cycle plays a pivotal role in regulating the galaxy formation and evolution process through cosmic time. As cosmic gas accretes in the dark matter halo potential well, it can cool and eventually form stars; massive stars explode as supernovae, enriching the surrounding interstellar medium with metals, that can potentially be ejected from the galaxy. Therefore key information regarding the star formation and enrichment history is encoded in the relation between stellar mass ($M_\star$) and gas metallicity ($Z_g$), the so-called mass-metallicity relation (MZR), making it effectively a standard ruler for galaxy evolution \citep[see][for a review]{maiolino:2018}.
%

Over the past 20 years, it has become possible to measure the MZR from $z\simeq 0$ \citep{tremonti:2004,kirby:2013,andrews:2013} up to the beginning of cosmic noon \citep[$z\simeq3$,][]{erb:2006,maiolino:2008,zahid:2011,sanders:2018,curti:2020}, mostly by using the oxygen to hydrogen abundance ratio [O/H] as a proxy for $Z_g$.
Locally, these observations have shown that $Z_g$ increases with stellar mass up to $M_\star\simeq 10^{10} \msun$, and beyond that point it flattens.
%
This can be understood as due to the fact that low-mass galaxies are prone to metal ejection due to their shallower potential well, while massive galaxies can retain most of the metals they have produced \citep{ferrara:2008}. In this view, galaxy evolution is guided by a \quotes{bathtub} equilibrium \citep{bouche:2010,dekel:2014} which determines the metallicity by balancing infall and outflows \citep{lilly:2013}.

Further, observations have highlighted that $Z_g$ and $M_\star$ are tightly connected with the star formation rate ($\rm SFR$) in the so-called fundamental mass-metallicity relation (FMR, \citealt{mannucci:2010}), which seems to be redshift-independent up to $z\simeq 3$ \citep{curti:2020}; other factors as galaxy size \citep{ellison:2008} and molecular gas content \citep{bothwell:2013} might also play some secondary role in connecting $Z_g$ to $M_\star$.

The MZR evolution at $z\gsim4$ has been predicted via semi-analytical models \citep{dayal:2013,zahid:2014,ucci:2023} or cosmological simulations \citep{torrey:2019,liu:2020,langan:2020,wilkins:2023, casavecchia:2024}. Most models predict almost no evolution or a very weak redshift dependence for $z>3$ \citep{ma:2016,kannan:2021,marszewski:2024}; some others suggest a breakdown of the relation at $z\gsim 9-12$ \citep[][]{pallottini:2014, sarmento:2018}.

Thanks to the exquisite spectroscopic capabilities of JWST, we can accurately infer $Z_g$ for galaxies well within the Epoch of Reionization (EoR, $z>6$). At present, surveys like JADES \citep{bunker:2023} and CEERS \citep{finkelstein:2023}, have provided metallicity measurements at $z\simeq 3-10$ \citep[][]{curti:2024} and $z\simeq 4-10$ \citep[][]{nakajima:2023}, for a combined sample of about 170 galaxies in the $5\times 10^6~\msun \lsim M_\star \lsim 10^{10}~\msun$ stellar mass range.

These JWST observations suggest that the MZR is already in place in the EoR, albeit downshifted by 0.5 dex with respect to the local one. The evolution in the $z\simeq 3-10$ range is very mild, as relatively low mass ($M_\star\simeq 10^{9-10}\msun$) galaxies can be chemically mature ($Z_g\simeq 0.3\, \zsun$) already at $z\simeq 6$. The MZR shows a r.m.s. dispersion $\sigma_Z = 0.3$ dex, and possibly a hint of flattening at $z\gsim 6$ \citep{curti:2024}.

Furthermore, multiple line detections in the same target can be used to constrain abundance patterns of individual elements \citep{kobayashi:2024, curti:2025}. Resolved observations are beginning to probe the presence of metallicity gradients up to $z\simeq 8$ \citep{venturi:2024}, which should provide deeper insights into the formation process of the galaxy and the history of the stellar mass buildup \citep[see][for an ALMA perspective]{vallini:2024}.
Observations targeting lensed fields \citep[][]{bezanson:2022,chemerynska:2024} are uncovering the metallicities of even the fainter galaxies \citep{chemerynska:2024_b}, which is key in order to explore the infall/outflow interplay due to stellar feedback in the less massive objects.
Importantly, using observations of lensed targets at $z\simeq 3-9$, \citet{morishita:2024} suggest that the 0.5 dex shift of normalization and the higher scattering\footnote{As far as we are aware, specific studies of the intrinsic scatter of the MZR relation are not present for JWST targets \citep[cfr. with][for $z=2-3$ scattering determinations]{strom:2022}, thus we take the observed scattering at face value.\label{footnote:face_value}} (0.3 dex, see also \citealt{heintz:2023}) w.r.t. the local MZR can be caused by the high level of burstiness of the SFR expected in high-$z$ galaxies \citep{pallottini:2023,sun:2023_b}.

Indeed, as we move from the local Universe to high $z$ \citep[see][for reviews]{madau:2014,dayal:2018,forster_schreiber:2020}, galaxies are expected to become more bursty, as a consequence of the increase of specific SFR \citep[sSFR, ][]{gonzalez:2010,stark:2013,smit:2014} combined with a decrease in size \citep{shibuya:2015}, which can make the stellar feedback more effective \citep{krumholz:2016} in maintaining a higher level of turbulence \citep{simons:2017,genzel:2017}.
%
With the recent availability of JWST observations \citep{borsani:2022,castellano:2022,finkelstein:2022_a,naidu:2022,treu:2022,adams:2023,atek:2023,donnan:2023,harikane:2023, santini:2023}, the time-varying stochastic SFR behavior (in short, flickering or burstiness), has been invoked as a possible mechanism \citep{mason:2023,mirocha:2023,shen:2023,sun:2023_b,kravtsov:2024} to explain the overabundance of bright galaxies at $z\gsim10$ \citep[see][for alternatives]{dekel:2023, ferrara:2023}.
However, some works predict that the r.m.s. amplitude of the SFR variability falls short of explaining the said overabundance \citep{pallottini:2023}. Thus, it is unclear if flickering alone can explain the phenomenon, particularly since the required high level of variability \citep{munoz:2023} is apparently not seen in the data \citep{ciesla:2024}.

Studies of the SFR flickering have experienced a recent surge, as stochastic SFR variations can have an impact on our ability to observe distant galaxies \citep{sun:2023} and affect the possibility of detecting Population III stars \citep{vikaeus:2022,katz:2023}. In addition, the flickering can modify the UV spectral slopes of high-$z$ galaxies and the escape fraction of ionizing photons \citep{gelli:2025}, ultimately controlling the cosmic reionization history \citep{davies:2016,nikolic:2024}.

Analyzing the amplitude and temporal variation of SFR flickering provides unique insights into the various feedback processes that regulate the formation and evolution of early galaxies \citep{pallottini:2023}.
On the one hand, high-amplitude and high-frequency flickering, on timescales shorter than the delay between SFR and SN explosions ($\simeq 20-40\,\myr$) are invoked \citep{gelli:2023, gelli:2024, gelli:2025_bursty} to explain the observed abrupt quenching of SFR in $M_\star\simeq 10^9\msun$ galaxies already at $z\simeq 7$ \citep{dome:2023, looser:2023, endsley:2025}.
On the other hand, the presence of disks observed in $M_\star\simeq 10^{9-10}\msun$ galaxies up to $z\simeq 7$ \citep{rowland:2024,fujimoto:2024} implies a limited level of stochasticity, as otherwise the disk would be disrupted \citep{kohandel:2024}.

In the present work, we aim to clarify the effects of SFR flickering and feedback regulation of high-$z$ galaxies by using the MZR as a standard ruler.
The paper is structured as follows; in Sec. \ref{sec:mzr_sim_and_obs} we present the \code{serra} cosmological radiation-hydrodynamic simulation suite and compare it with MZR observations from JADES, CEERS, and UNCOVER. To obtain a broad-brush interpretation of the complex, non-linear, effects shaping the MZR, in Sec. \ref{sec:analytical_model} we develop a minimal, single-zone model to give a physical interpretation to the observed MZR. In Sec. \ref{sec:model_discussion_and_caveats} we discuss the interpretation of different simulations suites by analyzing them in the light of the simplified view enabled by the minimal model (Sec. \ref{sec:sub:simulations_comparison}), also discussing its shortcomings (Sec. \ref{sec:sub:caveats}); conclusions are given in Sec. \ref{sec:conclusions}.

\section{The mass-metallicity relation at high redshift}\label{sec:mzr_sim_and_obs}

\begin{figure}
\centering
\includegraphics[width=0.49\textwidth]{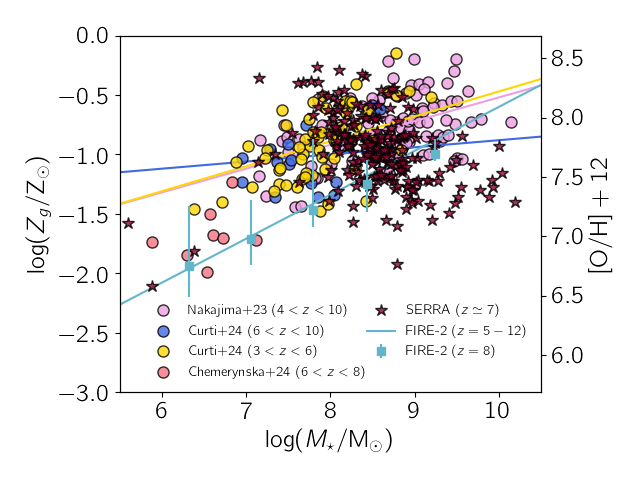}
\caption{The mass-metallicity relation (MZR) at high-redshift ($z$). We plot the gas metallicity ($Z_g$) as a function of stellar mass ($M_\star$) for the \code{serra} simulated galaxies \citep[][$z\simeq 7$]{pallottini:2022} and the JWST observations from CEERS \citet[][$z\simeq 4-10$]{nakajima:2023}, JADES \citep[][$z\simeq 3-6$ and $6-10$]{curti:2024}, and UNCOVER \citep[][$z\simeq 6-8$]{chemerynska:2024_b}.
To guide the eye, the observational fits from \citet{nakajima:2023} and \citet{curti:2024} are also reported as lines with the same colors of corresponding data.
As a reference, the $\log (Z_g/\zsun) = 0.37 \log(M_\star/\msun) - 4.3$ fit from \code{fire-2} simulations \citep[][$z=5-12$]{marszewski:2024} is shown, along with their binned data and scatter at $z=8$.
The right axis reports the logarithmic oxygen abundance [O/H] assuming a solar composition \citep{asplund:2009}, i.e. ${12+[{\rm O/H]_{\odot}}}= 8.69$.
\label{fig:MZR_obs_and_serra}
}
\end{figure}

To study the MZR at high-$z$, firstly we adopt the data from \code{SERRA}, a suite of cosmological zoom-in simulations that follows the evolution of $z\gsim 6$ galaxies \citep[][]{pallottini:2022}.
In \code{SERRA} we use the code \code{RAMSES} \citep{teyssier:2002} to evolve DM, gas, and stars. We enable the module \code{RAMSES-RT} \citep{rosdahl:2013} to track radiation on-the-fly by coupling it \citep{pallottini:2019,decataldo:2019} with \code{KROME} \citep{grassi:2014}, in order to account for non-equilibrium chemistry up to H$_{2}$ formation \citep{pallottini:2017_b}.
Stars are formed using a \citet{schmidt:1959}-\citet{kennicutt:1998} like relation based on H$_2$. Once formed, stars produce UV radiation that ionizes the gas, photo-dissociates H$_2$, and builds a radiation pressure in the gas. Massive stars generate stellar winds and explode as SNe. Depending on the kind of feedback, energy injection in the gas can be of thermal (subject to radiative losses) and/or turbulent (later dissipated) nature \citep{pallottini:2017}. In \code{SERRA}, metallicity is tracked as the sum of heavy elements.
Adopting a $\simeq 1.2\times 10^4 \msun$ ($\simeq 20\,{\rm pc}$) gas mass (spatial) resolution, with \code{music} \citep{hahn:2011} we initialize cosmological initial conditions\footnote{We adopt a $\Lambda$CDM model with vacuum, matter, and baryon densities in units of the critical density $\Omega_{\Lambda}= 0.692$, $\Omega_{m}= 0.308$, $\Omega_{b}= 0.0481$, normalized Hubble constant $h = \rm H_0/(100\, {\rm km}\,{\rm s}^{-1}\,{\rm Mpc}^{-1}) = 0.678$, spectral index $n=0.967$, and $\sigma_{8}=0.826$ \citep[][]{planck_collaboration:2014}.} generated at $z=100$, and follow the evolution of galaxies down to $z=6$ (and $z=4$ for a sub-sample, see \citealt{kohandel:2024}).

For this work, we select $\simeq 200$ central \code{serra} galaxies at $z=7.7$ with stellar mass $10^6\msun \lsim M_\star \lsim 10^{10} \msun$; we compute the gas metallicity as $Z_g = M_{g}^{Z}/M_g$ by considering the total gas ($M_g$) and metal ($M_Z$) mass contained within 3 stellar effective radii from the galaxy center, which is close to the gas half-mass radius\footnote{Different assumptions on the integration radius produce only small changes in the inferred metallicity.}.
Note that, for a more direct comparison with observation, we should compute line and continuum emission from \code{serra} galaxies; then, we could infer $M_\star$ from synthetic SED fitting and combine line emission to prepare metallicity calibrators. Such an approach would reduce potential mismatches due to uncertainties in using different calibrators \citep{curti:2017,chemerynska:2024_b}, e.g. to relax the single zone model assumption \citep[][]{marconi:2024} and correct for biases in the comparison \citep{cameron:2023}, also avoiding potential issues in the $M_\star$ determinations \citep{narayanan:2024}, which can hinder a robust determination of the MZR and its intrinsic scatter \citep{strom:2022}.

Without FIR data, metallicity can be typically inferred for galaxies showing emission lines either constraining the electron temperature \citep{vallini:2021, markov:2022, morishita:2024} or that can be combined to derive classical metallicity calibrators \citep[e.g.][]{curti:2020}. By definition, this is only a subset of star forming galaxies. It has to be noted that for some mini-quenched or post-starburst galaxies \citep{endsley:2025} metallicity estimates are also possible \citep{carniani:2025}.

While the fidelity of the comparison with observations might be improved by selecting from our model only star forming galaxies (and imposing a SFR lower limit to mimic a UV survey limit), the presence of lines for metallicity determination depends on assumptions such as the IMF type, the fraction of binaries \citep[e.g.][]{veraldi:2025}, and dust attenuation \citep{gelli:2023}. These choices add extra uncertainty to the models; future dedicated work is needed to explore this issue.

Also, while this kind of forward modeling approach is very powerful, also allowing the preparation of novel observational strategies \citep{zanella:2021,rizzo:2022}, here it hinders the possibility of comparing with most of the other models, as the MZR is usually computed integrating metal, gas, and stellar masses; thus we avoid it in the present work.

The prediction for \code{serra} are plotted in Fig. \ref{fig:MZR_obs_and_serra}, along with JWST data from \citet[][JADES sample, \citealt{bunker:2023}]{curti:2024}, \citet[][CEERS sample, \citealt{finkelstein:2023}]{nakajima:2023}, and \citet[][UNCOVER sample, \citealt{bezanson:2022}]{chemerynska:2024_b}, for a total of about 180 galaxies at $z\simeq 3-10$ with $10^6 \msun \lsim M_\star \lsim 10^{10} \msun$.
The bulk of \code{serra} galaxies is between $10^7\msun \lsim M_\star \lsim 10^9 \msun$; these galaxies have $10^{-1.5} \zsun \lsim Z_g \lsim 10^{-0.5}\zsun$, which is broadly consistent with the observed JWST galaxies.
However, at $10^7\msun \lsim M_\star \lsim 10^8 \msun$ some \code{serra} galaxies show $Z_g \simeq 0.25$ dex higher than the observed values; moreover, for $M_\star \simeq 10^{10} \msun$, $Z_g$ is on average lower than observed, up to extreme cases of $0.5$ dex. In general, \code{serra} galaxies show no clear MZR trend and their dispersion appears to be somewhat larger than the observed one.
Interestingly, with respect to \citet{nakajima:2023} and \citet{curti:2024}, the lensed galaxies from \citet{chemerynska:2024_b} seem to indicate lower values of $Z_g$ at $M_\star\lsim 10^7\msun$; the lensed data seems consistent with \code{SERRA}, but only a handful of simulated galaxies and observations are in that mass range.

As noted by \citet{curti:2024}, the lack of a clear trend and the large scatter in the \code{SERRA} relation resembles the behavior of their $6<z<10$ sample. However clear differences are present, such as the relatively metal-rich galaxies at $M_\star\simeq 10^{8} \msun$ and low metallicity galaxies at $M_\star\simeq 10^{10}\msun$.
Further, the lack of a clear trend might be partially due to galaxy sample selection, as most of the \code{SERRA} data has $10^8\msun \lsim M_\star\lsim 10^9\msun$ and only a small fraction of galaxies has $M_\star\simeq 10^7\msun$. Thus, the observed galaxy mass range is not uniformly covered by \code{serra} galaxies, whose sample is drawn from a collection of zoom-in simulations focusing on $M_\star \simeq 10^{10} \msun$ galaxies and their close environment. Indeed, some of the $M_\star\lsim 10^{9} \msun$ mass galaxies form in a highly pre-enriched environment, which explains their relatively high metallicities ($Z_g\simeq 10^{-0.5}$, see \citealt{gelli:2020}). Note that considering different redshift intervals\footnote{We have checked the \code{SERRA} data at $z=7\,,8,\,{\rm and}\, 9$.} yields qualitatively similar results.

Thus, in summary, the normalization of the MZR of \code{serra} is roughly similar to the observed galaxies, but no trend (flat MZR) is clearly visible \citep{rowland:2025} and some discrepancies are present, especially in the $M_\star \simeq 10^{10} \msun$ range.
Similar (or larger) tensions are present also when a comparison with other models is performed. These can be noted, e.g. from the fit obtained by the \code{fire-2} simulations \citep[][]{marszewski:2024}. \code{fire-2} data are close to the observed $Z_g$ at $M_\star\simeq 10^{9.5}\msun$, but the predicted slope of the MZR is different, thus the tension with \citet{nakajima:2023,curti:2024} increase with decreasing $M_\star$. Finally, \code{fire-2} galaxies present a much higher scatter\footnote{While \code{fire-2} adopts $\zsun=0.02$ as \code{SERRA}, the former simulation also track O abundance; \citet[][]{marszewski:2024} finds $[{\rm O/H}]_{\odot} +12 = 9.0$ for \code{FIRE-2} instead of the $[{\rm O/H}]_{\odot} +12 = 8.69$ \citep{asplund:2009} adopted here for the conversion of the observations. Adopting the $[{\rm O/H}]_{\odot} +12 = 9.0$ conversion for \code{FIRE-2} would make the agreement with observations better in the low mass regime, leaving unaltered the slope, scatter, and relative considerations.}, that increases with decreasing stellar mass up to $\simeq 1$ dex at $M_\star\simeq 5\times 10^{7}\msun$ (for a clearer comparison, see later Fig. \ref{fig:delta_vs_mstar}).

An in-depth comparison between observations and simulations is presented in \citet{curti:2024}, which, in addition to \code{serra}, considers the zoom-in simulations from \code{firstlight} \citep[][see \citealt{ceverino:2017} for the main paper]{langan:2020} and \code{FIRE} \citep[][see \citealt{hopkins:2014}]{ma:2016}, the IllustrisTNG cosmological simulations \citet[][see \citealt{pillepich:2018}]{torrey:2019}, and the \code{Astraeus} semi-analytical models \citep[][see \citealt{hutter:2021}]{ucci:2023}; as detailed in \citet{curti:2024}, most simulations seem to reasonably match the $Z_g$ data at $\simeq 10^9 \msun$ but have steeper slopes and underpredict the metallicities at lower $M_\star$.

Interestingly, some of the models \citep[e.g.][]{marszewski:2024,cueto:2024} seem to be consistent with what was observed by \citet{chemerynska:2024_b}, which however have only eight targets, thus it is unclear if there is a change of slope of the MZR for $M_\star\lsim 10^{7}\msun$.
Indeed, the observational determination of the slope of the MZR seems still uncertain, e.g. it seems to change with redshift, as indicated by the different slopes resulting from the separate analysis of \citet{curti:2024} for the $3< z<6$ and $6< z<10$ samples.

\section{A minimal physical model to explain the MZR}\label{sec:analytical_model}

Given the very contrived match between simulations and observations, it seems worthwhile to step back and try to understand the trend and dispersion of the MZR using basic physical models. The advantage of the approach consists of simplifying the complexity of numerical simulation while retaining the most important physical processes shaping the MZR. 

To this aim, we devise a minimal, physical model describing the evolution of DM ($M_{dm}$), gas ($M_g$), star ($M_\star$), and gas-phase metal ($M_g^Z$) mass of a galaxy, following \citet{dayal:2013}.
As in \citet{furlanetto:2022}, we introduce an explicit feedback delay resulting in a modulation of the star formation history (SFH).

\subsection{Model setup}\label{sec:sub:model_flux}

We assume that the DM halo increases as a result of cosmological accretion, which on average can be written as in \citet[][see eq. 23 therein]{correa:2015_b}
\begin{subequations}\label{eqs:mass_flux}
\begin{equation}\label{eq:mhalo_growth}
\dot{M}_{dm} = 71.6 \left(\frac{M_{dm}}{10^{12} \msun}\right) \left(\frac{0.7}{h}\right)\, E(z)\,\msunyr\,,
\end{equation}
where $E(z) = [-0.24 + 0.75 (1+z)]\sqrt{\Omega_m(1+z)^3 + \Omega_\Lambda}$.
We allow stars to form (eq. \ref{eq:sfr}) and include stellar feedback via SNe, exploding after a delay time $\tdelay$. SNe cause mass outflows with a rate $\dot{M}_{out}$ (eq. \ref{eq:outflow}), enrich the gas of metals with a yield $y$, and have a return fraction $R$.
Thus, the stellar mass increases because of SFR and decreases after $\tdelay$ because of SN explosions:
\begin{equation}\label{eq:evol_star}
\dot{M}_\star = {\rm SFR}(t) - R\,{\rm SFR}(t-\tdelay)\,,
\end{equation}
which for $t_d=0$ (no-SN-delay) would give an instantaneous recycling approximation as in \citet{dayal:2013}.

The gas mass increases because of cosmic accretion and return from stars, and it decreases because of the SFR process and SN-driven outflows:
\begin{equation}\label{eq:evol_gas}
\dot{M}_g = f_b\dot{M}_{dm}(t) + R\,{\rm SFR}(t-\tdelay) - [{\rm SFR}(t) + \dot{M}_{out}(t-\tdelay) ]\,,
\end{equation}
where $f_b = {\Omega_b}/{\Omega_m}$ is the cosmological baryion fraction.
Finally, the gas metal content increases because of the SN yields while it decreases because of astration and outflows:
\begin{equation}
\dot{M}_g^Z = f_b Z_{in}\dot{M}_{dm}(t) + y\,{\rm SFR}(t-\tdelay) - Z_g [{\rm SFR}(t) +\dot{M}_{out}(t-\tdelay)]\,, \label{eq:evol_metal_gas}
\end{equation}
where $Z_g = M_g^{Z}/M_g$ is the gas metallicity, and we include the possibility for the cosmic gas to be already enriched at $Z_{in}$, which we set to $Z_{in}=10^{-7}\zsun$, i.e. virtually metal free.
Note that in principle the enrichment ($y \mathrm{SFR}(t-\tdelay)$ in eq. \ref{eq:evol_metal_gas}) and outflow ($\dot{M}_{out}(t-\tdelay)$ in eq. \ref{eq:evol_gas}) time delays might be different, as the former is linked to stellar evolution, while the latter describes mechanical feedback that is powering galaxy outflows.
Similarly, the outflow gas metallicity might be different from the ISM one, while this possibility cannot be included in a single-zone model.
Such effects are sometimes considered in more complex semi-analytic models accounting for a two-phase medium \citep[cfr.][]{sommerville:2015, mutch:2016}. For the sake of simplicity, here we do not allow for this additional (and uncertain) factor.
\end{subequations}

To solve the system in eq.s \ref{eqs:mass_flux}, we need to specify a functional form for $\dot{M}_{out}$ and SFR.
We take the outflow rate as in \citet[][]{muratov:2015}:
\begin{subequations}\label{eq:sfr_and_outflow}
\begin{equation}\label{eq:outflow}
\dot{M}_{out} = 3.6\,\epsilon_{\rm SN}\, {\rm SFR} \left(\frac{M_\star}{10^{10}\msun}\right)^{-0.35}\msunyr\,,
\end{equation}
with $\epsilon_{\rm SN}$ an efficiency parameter that we use to calibrate the mass loading factor; $\epsilon_{\rm SN}=1$ correspond to the results from \citet[\code{fire} simulations, \citealt{hopkins:2014}][]{muratov:2015}. Note that $\epsilon_{\rm SN}=1$ yields a loading factor consistent with what \citet{herrera_camus:2021} observes for HZ4 \citep[but see][]{parlanti:2024}. Further, using $\epsilon_{\rm SN}=1$ in eq. \ref{eq:sfr_and_outflow} would match the results from \citet{pizzati:2023} for $z\simeq 5-6$ ALPINE galaxies \citep{le_fevre:2019} which feature a spatially extended gas component \citep{fujimoto:2020}. However, $\epsilon_{\rm SN}\simeq 1/3$ is needed to match the loading factor inferred from $z\simeq 3-9$ JADES galaxies \citep[][]{carniani:2024}.
We note that simulated outflow loading factors \citep[][]{muratov:2015} are typically computed in a different way than inferred from observations \citep[e.g.][]{pizzati:2020,carniani:2024}. Thus, some care should be taken in the comparison, and our conclusions should be taken with caution as they are based on simplified assumptions.

We assume that the SFR is proportional to the gas mass:
\begin{equation}\label{eq:sfr}
{\rm SFR} = \frac{M_g}{t_\star}\,,
\end{equation}
\end{subequations}
and we select a constant $t_\star = 1\,\rm Gyr$ as our fiducial star formation time scale\footnote{In the model, galaxy evolution is self-regulated by feedback, thus the sensitivity of the results from $t_\star$ is limited in our case, and null in the $\tdelay=0$ case with constant loading factors for infall and outflow \citep{lilly:2013}. \textit{A posteriori}, we note that the fiducial $t_\star = 1\,\rm Gyr$ gives specific star formation rates ${\rm sSFR} = {\rm SFR}/M_\star$ which are consistent with \code{SERRA} galaxies, i.e. $\simeq {\rm sSFR}\simeq 100-1 \,\rm Gyr^{-1}$ for galaxies with masses $M_\star=10^6 - 10^{10} \msun$.}. The choice of $t_\star$ is roughly consistent with the upper limits of the depletion times reported by \citet{dessauges_zavadsky:2020} for ALPINE galaxies, and it is discussed in more detail in Sec. \ref{sec:sub:caveats} (see also App. \ref{app:sec:modify_sfr}).

To complete the model, we assume that halos can form stars after they reach a virial temperature for atomic cooling to be effective ($10^4\,\rm K$); similarly to \code{serra}, we assume return fraction and yields appropriate for a \citet{kroupa:2001} IMF and adopting \citet{bertelli:1994} tracks with a $Z_\star=\zsun$ stellar population\footnote{Changing the metallicity of the stellar population does modifies the yield and return fraction, but gives only minor modifications to the upcoming results.}, which gives $y\simeq 0.0228$ and $R\simeq 0.3242$.

The system (eq.s \ref{eqs:mass_flux} and \ref{eq:sfr_and_outflow}) is solved along with $z=z(t)$ using an embedded 5th/6th order Runge-Kutta method \citep{fehlberg:1968} with adaptive timestep, which is selected in order to have a fractional (absolute) precision of at least $\leq 10^{-5}$ ($\leq 10^2 \msun$) for $M_{dm}$, $M_\star$, $M_g$, and $M_{g}^{Z}$. The delayed SFR and stellar mass used to evaluate the stellar feedback are computed using a 7-point time stencil.

\subsection{Overview of the minimal physical model}\label{sec:model_overview}

\begin{figure}
\centering
\includegraphics[width=0.49\textwidth]{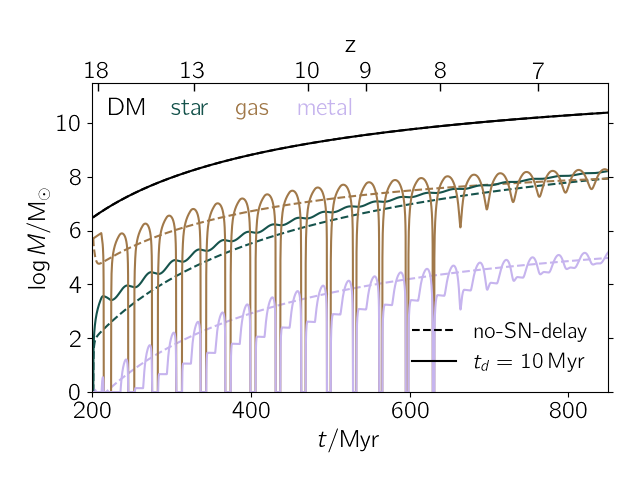}
\caption{
Example of the impact of SFR burstiness on the evolution of a galaxy hosted in a $M_{dm} = 3\times 10^{10}\msun$ DM halo at $z=6$ using the minimal physical model described in Sec. \ref{sec:sub:model_flux} (see in particular eq.s \ref{eqs:mass_flux}).
The time evolution of the dark matter (DM, black), star (green), gas (brown), and metal (pink) mass ($M$) content is shown with solid (dashed) lines for a SN delay time $\tdelay =10\,\myr$ ($\tdelay = 0$, i.e. no delay).
The upper axis shows the redshift $z$ corresponding to cosmic time $t$.
\label{fig:analytical_model_examples}
}
\end{figure}

\begin{figure}
\centering
\includegraphics[width=0.49\textwidth]{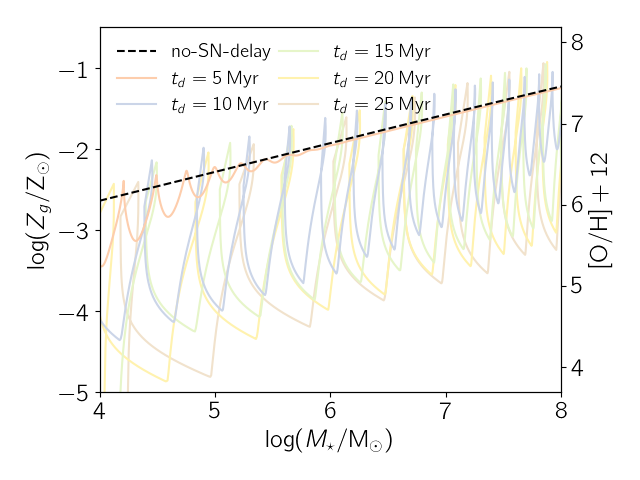}
\caption{
An example of the impact of SFR burstiness on the MZR for a system with $M_{dm} = 3\times 10^{10}\msun$ at $z=6$ evolved with our minimal physical model.
Different lines indicate different delay times for the SN feedback, as indicated in the legend.
Note that the model results have been resampled in $0.2\,\myr$ steps and we do not consider the evolutionary phases when the galaxies have a negligible amount of gas ($M_g\leq 10^{-3}M_{dm}$), which allows to have a well defined $Z_g$ even in periods when no gas and metals are present in the galaxy (see Fig. \ref{fig:analytical_model_examples}).
For the effect of a coarser time sampling, see App. \ref{app:sec:coraser_dt}, in particular Fig. \ref{fig:app:analytical_model_MZR_examples_coarse}.
\label{fig:analytical_model_MZR_examples}
}
\end{figure}

The delay between SFR and feedback $\tdelay$ and the amplitude of the outflow ($\epsilon_{\rm SN}$) regulate the burstiness of the galaxy in our single zone model \citep[cfr.][]{furlanetto:2022}.
To understand the impact of burstiness on galaxy evolution, it is instructive to consider a halo with mass $M_{dm} = 3\times 10^{10} \msun$ at $z=6$ and set the outflow efficiency to the standard value of $\epsilon_{\rm SN}=1$.

In Fig. \ref{fig:analytical_model_examples} we compare the evolution of the stellar (green), gas (brown), and metal (pink) mass content for $\tdelay = 0$ (no-SN-delay, dashed lines) and $\tdelay = 10\, \rm Myr$ (solid lines).
By construction, the halo mass growth (eq. \ref{eq:mhalo_growth}) is unaffected by $\tdelay$, thus the two models follow the same DM evolution (black line), which smoothly grows from $M_{dm} \simeq 10^7 \msun$ at $z \simeq 18$, when the universe was $t\simeq 200\,\myr$ old.
At that time, for $\tdelay=0$ the gas and stars has $\simeq 10^5 \msun$ and $\simeq 10^2 \msun$, respectively, and both grow reaching approximately $\simeq 5\times 10^7 \msun$ at $t\simeq 800\,\myr$, with $M_\star/M_{dm}\simeq 5\times 10^{-3}$ broadly consistent with what expected from abundance matching models \citep[e.g.][]{behroozi:2013_c}.

In the case with $\tdelay = 10\, \rm Myr$, the secular trends are qualitatively similar, but the system experiences a series of bursts on time scales of $\simeq 20\, {\rm Myr}=2\,\tdelay$.
Let us focus on a single burst cycle around $z\simeq 10$, when the halo has $M_{dm}\simeq 10^{9}\msun$ and $M_\star\simeq 10^6\msun$. In the first half period of the cycle ($0-\tdelay$), cosmic accretion quickly replenish the gas (gas infall rate is $\simeq 0.5\, \msunyr$ from eq. \ref{eq:mhalo_growth}), reaching $M_g\simeq 10^7\msun$ in a few Myr. Stars form at a modest rate ${\rm SFR}\simeq 10^{-2}\msunyr$ (eq. \ref{eq:sfr}) under feedback-free conditions \citep[cfr.][]{dekel:2023}. After $\tdelay$, the feedback kicks in, first enriching the galaxy and eventually ($\simeq 2\,\tdelay$) becoming powerful enough to expel both gas (gas outflow is $\simeq 1 \msunyr$ using eq. \ref{eq:outflow}) and metals \citep[cfr.][]{ferrara:2023}.

Note that the delay effectively produces a sequence of mini-quenching periods \citep[cfr.][]{looser:2023}, which are commonly inferred in faint galaxies \citep{endsley:2025}, and can be explained as a consequence of internal feedback \citep{gelli:2025_bursty}. However, in our minimal model, outflows are the only mechanism regulating the star formation, and quenching due to gas complete depletion which might be a too extreme description of the actual physical evolution \citep[but see][]{gelli:2024}. 
Further, in the example, the impact of SFR flickering is limited already after $z\simeq 15$, when the galaxy has $M_\star \simeq 10^{5} \msun$ or $10\times$ lower if no delay is considered. At $z\simeq 6$, the final stellar mass is $M_\star \simeq 10^7 \msun$ within a factor of 2 for the two models.
For the gas and metals, the model with SN delay starts to converge to the no-SN-delay case at $z\simeq 8$, when $M_\star \simeq 10^7 \msun$; oscillations for the gas and metals are still present up to $z\simeq 7$, but they are milder.

As in \citet{pallottini:2023}, it is convenient to quantify the burstiness of the SFR by defining the variable $\delta$ expressing the log of the stochastic SFR variation with respect to its mean (in short: flickering)
\begin{subequations}\label{eq:def_flickering}
\begin{equation}
\delta \equiv \log \frac{\rm SFR}{\langle{\rm SFR}\rangle}\,,
\end{equation}
where $\log \langle{\rm SFR}/\msunyr\rangle$ is a second order polynomial fit in the time variable $t$. The low order fit is needed to factor out the smooth behavior of the functions without removing time modulations \citep[see][for alternative methods]{leja:2019,montero:2021}. As in \citet{pallottini:2023}, we use a $2\,\myr$ time scale binning when evaluating SFR in eq. \ref{eq:def_flickering}, while no explicit time averaging is needed for $\langle{\rm SFR}\rangle$, since it is implicitly done via the polynomial fit \citep[cfr.][]{sun:2023_b}.

Similarly to the flickering, we define the analog variable for the gas metallicity
\begin{equation}
\delta_{Z} \equiv \log \frac{Z_g}{\langle Z_g\rangle}\,,
\end{equation}
\end{subequations}
where $\log \langle Z_g /\zsun\rangle$ is a second order polynomial fit in the variable $\log M_\star/\msun$. For both $\delta$ and $\delta_Z$ we can define the typical variation as the root mean square (r.m.s.) deviation, i.e. $\sigmasfr$ and $\sigmaz$, respectively.

Adopting eq.s \ref{eq:def_flickering}, we find that the $\tdelay = 10\, \rm Myr$ case has a flickering $\sigmasfr\simeq 0.3$ before the two models start to converge, which is slightly higher than the $\sigmasfr\simeq 0.24$ obtained for \code{SERRA} galaxies \citep{pallottini:2023}.
For different $\tdelay$ the qualitative behavior is very similar, as the model is off-balance and oscillates around the no-SN-delay case, which is similar to a bathtub solution \citep{bouche:2010,dekel:2014}. For shorter (longer) $\tdelay$, $\sigmasfr$ can be smaller (higher), as a galaxy with a smaller (higher) $M_\star$ can be off-balanced.

The balance between infall and outflow rates is critical since the dynamical system (eq.s \ref{eqs:mass_flux}) modeling the galaxy evolution gives rise to exponentially increasing/decreasing trends; introducing a $\tdelay\neq 0$ can efficiently break the balance, causing cycles of complete gas and metal depletion.

To explore the MZR variations induced by the flickering, in Fig. \ref{fig:analytical_model_MZR_examples} we plot the tracks of the selected $M_{dm} = 3\times 10^{10}\msun$ at $z=6$ with $t_d$ from 5 to 25 Myr.
Because of the longer $\tdelay$, the SFR becomes increasingly bursty. Specifically, for the selected model the flickering ranges from $\sigmasfr\simeq 0.1$ to $\sigmasfr\simeq 0.6$ for $\tdelay=5$ Myr to $\tdelay=25$ Myr, respectively.

The MZR modulations induced by the flickering are highly non-linear.
With $\tdelay=5-10\,\myr$, the MZR dispersion is negligible-small ($\sigmaz\simeq 0.05-0.2$), as the flickering can affect galaxies only up to $M_\star\simeq 10^{5-7}\msun$.
For $\tdelay=15\,\myr$, the full $M_\star$ range experience flickering with a $\sigmasfr\simeq 0.5$, yielding a high $\sigmaz\simeq 0.8$.
Cases with $\tdelay>15$ Myr do not show a higher stochasticity, as an increasing delay only regulates which $M_\star$ can be affected by the feedback unbalance\footnote{Note that a higher/lower level of flickering can be induced by changing the infall/outflow rates, as shown later in Sec. \ref{sec:impact_MZR}.}. Qualitatively, the behavior is similar to higher $M_{dm}$, but the convergence to the bathtub solution is faster.
For the effect of a coarse time sampling, see App. \ref{app:sec:coraser_dt} (in particular Fig. \ref{fig:app:analytical_model_MZR_examples_coarse}).

In summary, a feedback delay induces a stochastic, bursty star formation behavior that becomes progressively more enhanced as $t_d$ is increased. Such SFR flickering corresponds to analog $Z_g$ fluctuations, which act to decrease the mean metallicity as gas and metals are effectively ejected from the galaxy by outflows. 
%

\subsection{The impact of a stochastic SFR on the MZR}\label{sec:impact_MZR}

\begin{figure*}
\centering
\includegraphics[width=0.49\textwidth]{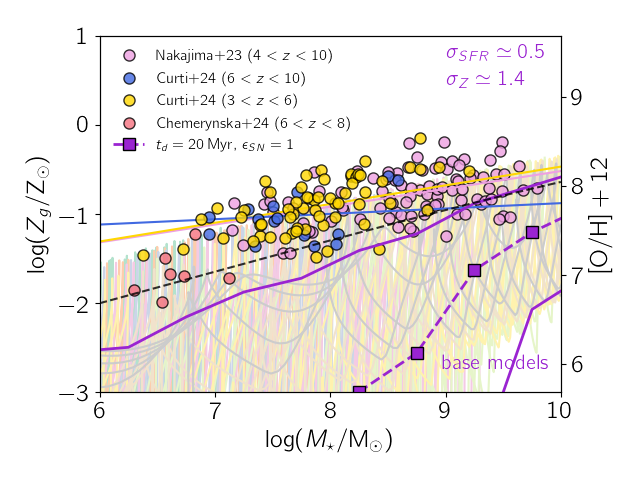}
\includegraphics[width=0.49\textwidth]{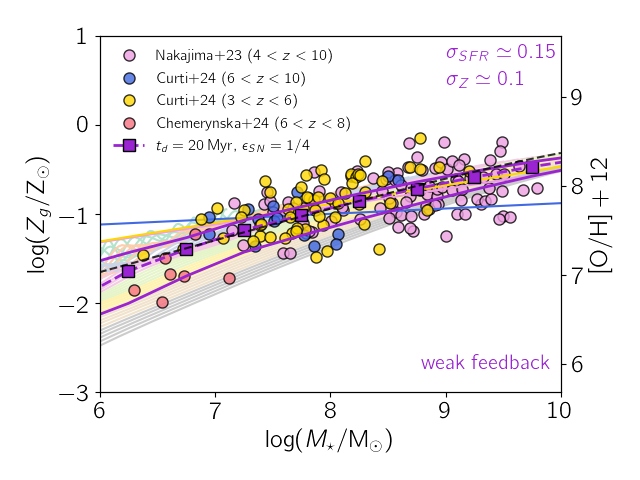}
\includegraphics[width=0.49\textwidth]{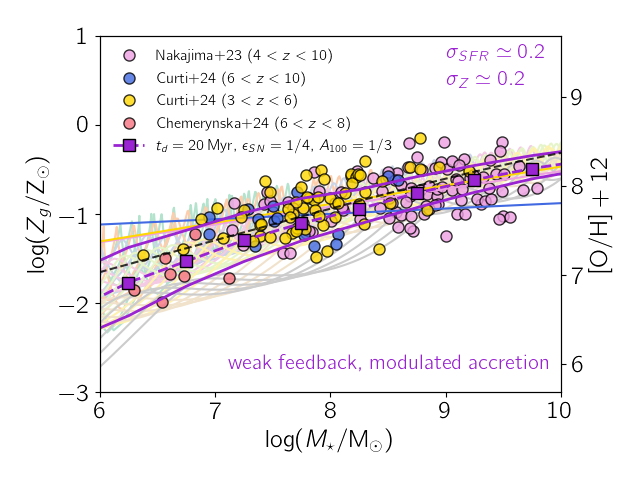}
\includegraphics[width=0.49\textwidth]{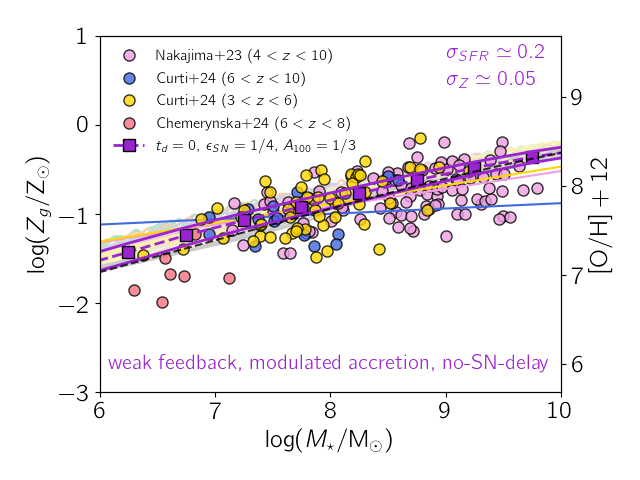}
\caption{Predicted MZR relation for different sets of minimal models with $10^{9}\msun \leq M_{dm} \leq 10^{13}\msun$ halos at $z=6$.
In the \textbf{upper left} panel we report the \textit{base models} with $\epsilon_{SN}$ and $\tdelay=20\,\myr$ (see Sec. \ref{sec:model_overview}).
In the \textbf{upper right} panel we modify the base model by reducing the efficiency of the feedback ($\epsilon_{\rm SN}=1/4$ in eq. \ref{eq:outflow}).
In the \textbf{lower left} panel we consider both weak feedback and a cosmic accretion that is modulated on a time scale of $100\,\myr$ ($\acosmo = 1/3$ in eq. \ref{eq:mhalo_growth_modified}).
In the \textbf{lower left} panel we show the effect of turning off the SN-delay for the weak feedback models with modulated cosmic accretion.
In each panel, pastel lines are the individual tracks sampled every $2\,\myr$; their median and $16\%-84\%$ dispersion are plotted with violet dashed and solid lines, respectively; the dashed black line is the control case with no-SN-delay ($\tdelay=0$) and no modulated accretion ($\acosmo = 0$).
For each model, we report the value values of the SFR flickering ($\sigmasfr$) and the MZR dispersion ($\sigmaz$, see Fig. \ref{fig:delta_vs_mstar}).
As in Fig. \ref{fig:MZR_obs_and_serra}, we show JWST data and fits \citep{nakajima:2023,curti:2024,chemerynska:2024_b}.
\label{fig:MZR_flickering_impact}
}
\end{figure*}

\begin{figure}
\centering
\includegraphics[width=0.49\textwidth]{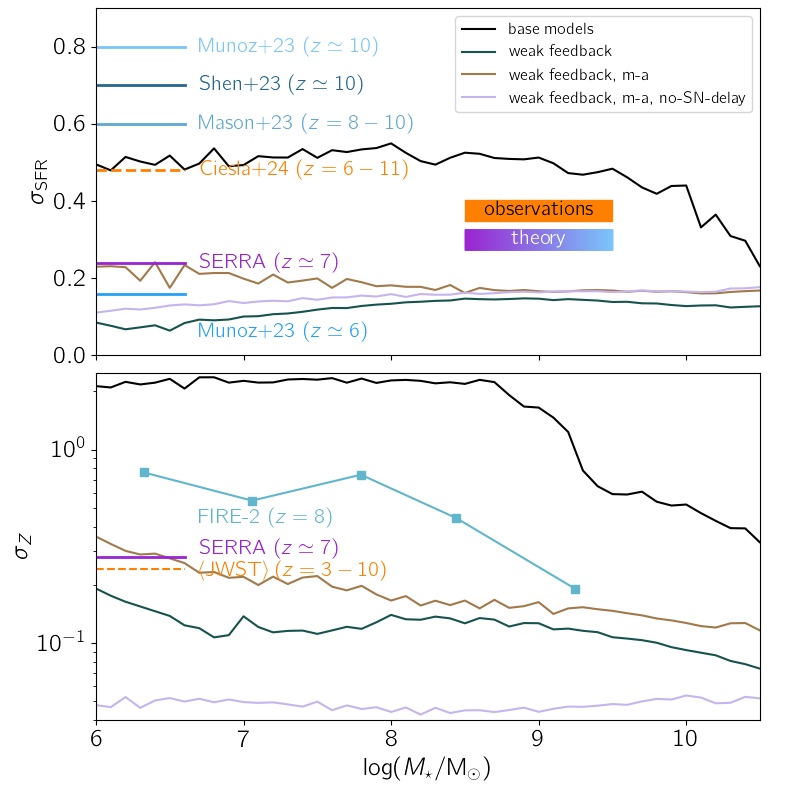}
\caption{SFR ($\sigmasfr$, upper panel) and gas metallicity ($\sigmaz$, lower panel) r.m.s. variations as a function of stellar mass for the sets of models presented in Fig. \ref{fig:MZR_flickering_impact} (see eq.s \ref{eq:def_flickering} for the definitions).
As a reference, we report the \code{SERRA} average $\sigmasfr$ \citep{pallottini:2023} and $\sigmaz$ (this work), the $\sigmasfr$ from \citet{ciesla:2024, mason:2023,munoz:2023, shen:2023}, and the $\sigmaz$ for \code{FIRE-2} at $z=8.0$ \citep{marszewski:2024} and JWST data.
Note that $\langle \rm JWST \rangle$ is the r.m.s. of the observations after subtraction of the fit to the MZR in the different data sets in \citet{nakajima:2023} and \citet{curti:2024}; the thickness of the shaded area encloses the min/max of the r.m.s. of the different data sets.
\label{fig:delta_vs_mstar}
}
\end{figure}

We run sets of models that at $z=6$ have a $M_{dm}$ from $10^{9}\msun $ to $10^{13}\msun$ with a logarithmic binning of $0.1$ dex, in order to cover the mass range $10^5 \msun \lsim M_\star \lsim 5\times 10^{10}\msun$.
We analyze different variations of the minimal models, reporting the results in different panels of Fig. \ref{fig:MZR_flickering_impact}, where they are compared with data from JWST observations \citep[][]{nakajima:2023,curti:2024,chemerynska:2024_b}.

\subsubsection*{Base models}

Let us start by discussing the \textit{base models} analyzed in Sec. \ref{sec:analytical_model}, i.e. with $\epsilon_{\rm SN}=1$, consistent with \code{fire} simulations \citep{muratov:2015}, and selecting $\tdelay=20\,\myr$, so that most of the explored stellar mass range is expected to have a stochastic SFR.

In the upper left panel of Fig. \ref{fig:MZR_flickering_impact} we report both the individual tracks of the base models and their $M_\star$ average trends.
For the base models, the median case falls significantly below the observed average MZR. Also, the dispersion is much larger than seen in JWST data, which can be barely matched by the upper envelope of the predicted curves. 
The base models have a $\sigmasfr\simeq0.5$ flickering, that induces a $\sigmaz\gsim 1$ almost in the entire stellar mass range ($M_\star\lsim 10^9\msun$, Fig. \ref{fig:delta_vs_mstar}).
The flickering is due to cycles of infall/outflow which determine an oscillatory behavior of the evolution around the $\tdelay=0$ control case, which is largely independent of the hosting $M_{dm}$.
The $Z_g$-$M_\star$ normalization is determined by the strength of the outflows, which completely devoids the galaxy of gas and metals, not allowing the metallicity to rise to the observed level. Even for $\tdelay=0$ the outflow seems too extreme, as only part of the lensed galaxy data can be matched \citep{chemerynska:2024_b}.

In summary, the efficient feedback ($\epsilon_{\rm SN}=1$) combined with a $\tdelay=20\,\myr$ generates a SFR flickering of $\sigmasfr\simeq 0.5$, inducing MZR variations of $\sigmaz\simeq 1.4$ resulting in a \quotes{chemical chaos} that is not observed. Barring the option that observations are tracing only the tip of the $Z_g$-$M_\star$ relation, in order to explain the observed MZR we modify the base models as follows.

\subsubsection*{Weak feedback}

First, we reduce the impact of stellar feedback by decreasing the outflow rate from \citet{muratov:2015} (eq. \ref{eq:outflow}), i.e. setting $\epsilon_{\rm SN}=1/4$; such a loading factor is closer to what \citet{carniani:2024} infer for $10^{7.5}\lsim M_\star \lsim 10^9\msun$ JADES galaxies.
As shown in the upper right panel of Fig. \ref{fig:MZR_flickering_impact}, this modification increases the average metallicity of the modeled galaxies, matching the observed mean trends in the whole $10^6\msun \lsim M_\star \lsim 10^{10}\msun$ mass range. Compared to the $\tdelay=0$ control case, it is evident that the outflow efficiency is the main driver of the mean $Z_g$-$M_\star$ trend.

Note that, reducing the outflow efficiency while maintaining a fixed $\tdelay=20\,\myr$ results in $\sigmaz\simeq 0.1$ modulation of the MZR, i.e. too low with respect to the observed ($\sigmaz\simeq 0.25$, see Fig. \ref{fig:delta_vs_mstar}).
This effect is induced by the suppression of the amplitude of the SFR flickering, especially at the high mass end.

Indeed the models show a small flickering, with $\sigmasfr\simeq0.15$ for $M_\star\lsim 10^8\msun$, and even slightly lower values for higher stellar masses (Fig. \ref{fig:delta_vs_mstar}).
As discussed in \citet{pallottini:2023}, $M_\star \lsim 10^9\msun$ galaxies show SFR fluctuations with characteristic periods of $\simeq 20-40\, \myr$, which correspond to the typical evolutionary times of SNe.
Such flickering timescale increases to $\simeq 80-150\, \myr$ in more massive galaxies which are more sensitive to mergers and accretion rate modulation. The latter effects are not accounted for in our minimal model (eq. \ref{eq:mhalo_growth}). The next step is then to heuristically incorporate them into our treatment.

\subsubsection*{Weak feedback, modulated accretion}

To incorporate the extra stochastic behavior expected from variation in the cosmic accretion/merger in a simple manner, we impose a sinusoidal modulation on a $100\, \myr$ timescale
\begin{equation}\label{eq:mhalo_growth_modified}
\dot{M}_{dm}^{mod}= \dot{M}_{dm} \left[1 + \acosmo\sin\left(\frac{2 \pi t }{100 \,\myr}\right)\right]\,,
\end{equation}
with $\dot{M}_{dm}$ being the average cosmic accretion from eq. \ref{eq:mhalo_growth} and with $\acosmo$ scaling the amplitude of the modulation\footnote{Note that small halos should be relatively less affected by modulation on long timescales ($\simeq 100\, \myr$), since the hosted galaxies have shorter SFH. However, cosmological accretion variations can be coherently combined with the SN delay to give a higher $\sigmasfr$ also in the low mass range.}.
For different galaxies, a phase shift is implicitly included because star formation starts when the hosting halo mass is larger than the atomic cooling halo threshold. While our inclusion of the accretion modulation is effective, it is very crude. For a more refined treatment see \citet[][]{sun:2024}, which adopts a modulation based on random extractions of variation of the accretion histories by using a pre-defined power spectral density.

As feedback determines the mean trend of the MZR, we keep the reduced outflow efficiency ($\epsilon_{\rm SN}=1/4$ in eq. \ref{eq:outflow}) and compute the evolution of a minimal model by setting the amplitude of the oscillation to $\acosmo=1/3$, which should yield a scatter below the 0.3 dex dispersion expected for the distribution of growth rates of DM halos \citep{rodriguez_puebla:2016, ren:2019, mirocha:2021}.
The results from these modified models are shown in the lower left panel of Fig. \ref{fig:MZR_flickering_impact}: both the mean trend and scatter of the observed MZR are recovered (Fig. \ref{fig:delta_vs_mstar}).
Such a matching set of modified models yields a roughly constant $\sigmasfr\simeq 0.2$ and $\sigmaz\simeq 0.3$, satisfactorily matching the JWST data.

\subsubsection*{Weak feedback, modulated accretion, no-SN-delay}

As a final check, in the lower right panel of Fig. \ref{fig:MZR_flickering_impact} we show the effect of removing the SN-delay from the best matching model just discussed.
While the mean MZR trend is recovered, the metallicity dispersion ($\sigmaz\simeq 0.05$) is much smaller than observed ($\sigmaz\simeq 0.25$). Also, the median of the models fails to reproduce the relation for most of the low-mass lensed galaxies \citep{chemerynska:2024_b}.

In the no-SN-delay case, the SFR flickering is induced only by modulation of the cosmic accretion, which fails to unbalance the system through the ejection of a substantial fraction of the contained gas.
With the selected $\acosmo=1/3$ and $\tdelay=0$, the model reach $\sigmasfr\simeq0.1$ for $M_\star\lsim 10^8\msun$, while the flickering gets to $\sigmasfr\simeq0.2$ only for $M_\star\simeq 5\times 10^{8}\msun$ galaxies.

Compared to the control case ($\acosmo=0$), the average $Z_g$ is slightly higher; practically, SFR flickering induced by the accretion modulation can increase $Z_g$ since a higher gas mass can be converted into stars without efficiently ejecting the gas via outflows, i.e. the opposite outcome with respect to SFR flickering induced via SN-delay.
Interestingly, despite the relatively high SFR flickering (average of $\sigmasfr\simeq 0.2$), the metallicity variation is the lowest of all considered cases ($\sigmaz\simeq 0.05$): only a delayed SN feedback can efficiently off-balance the system and induce modulations on both SFR and $Z_g$.

We note that the \code{serra} cosmological simulations show a relatively modest SFR flickering \citep[$\sigmasfr\simeq 0.24$,][]{pallottini:2023}, and close to the expectations for $z\simeq 6$ galaxies \citep{munoz:2023,ciesla:2024}. Yet, given the sharp sensitivity of $\sigmaz$ to SFR variations, such time variability results in a $\sigmaz$ larger than observed. 
To summarize, our analysis shows that: (i) weak feedback, and (ii) relatively mild stochasticity given by (iii) SN-delay combined with a long-term accretion rate modulation are critical to simultaneously match both the mean and dispersion of the observed MZR relation in the entire stellar mass range of galaxies sampled by JWST at $z=3-10$.

Interestingly, we note that invoking stochasticity to explain the overabundance of bright JWST galaxies requires $\sigmasfr>0.5$ (more specifically, $\sigmasfr\simeq 0.8$, $\sigmasfr\simeq 0.7$, and $\sigmasfr\simeq 0.6$ according to \citealt[][]{munoz:2023}, \citealt[][]{shen:2023}, and \citealt[][]{mason:2023}, respectively; see Fig. \ref{fig:delta_vs_mstar}).
However, such extreme flickering amplitudes are (i) higher than that recovered from SED fitting allowing for SFH variability \citep[$\sigmasfr\lsim 0.5$,][]{ciesla:2024}, and (ii) would induce a chemical chaos ($\sigmaz\gsim 1.4$) that is not present in the observed MZR at $z=3-10$ ($\sigmaz\simeq 0.25$ from \citealt{nakajima:2023, curti:2024,chemerynska:2024_b} data, $\sigmaz\sim 0.3$ reported by \citealt{heintz:2023, morishita:2024}).
Indeed, adopting the \code{FIRE-2} model, \citet{sun:2023_b} report the overabundance of bright super-early galaxies can be matched;  \citet{marszewski:2024} find a MZR dispersion at $M_\star\lsim 10^{8.5}\msun$ which is roughly twice the average $\sigma_Z$ observed JWST.
The drop of $\sigma_Z$ from \code{FIRE-2} at $M_\star\gsim 10^{8.5}\msun$ might be caused by the low number of galaxies in that mass bin (see Fig. 1 from \citealt{{marszewski:2024}}) or a reduction of the flickering at higher stellar masses (compare with Fig. 13 from \citealt{kravtsov:2024}).

Finally, note that \citet{katz:2024} reports that galaxies dominated by strong nebular continuum (Balmer jump galaxies) can have UV flux increases that might help in reconciling the surprising abundance of bright high-$z$ galaxies; however, the incidence of Balmer jump galaxies is unconstrained at $z\gsim 10$ and the magnitude of the effect is relatively small ($\simeq 0.28$ dex in terms of $\sigmasfr$): even if all the system were to be Balmer jump galaxies, high flickering (or a different mechanism) would still need to be invoked to explain the overabundance of bright super-early galaxies.

\section{Discussion}\label{sec:model_discussion_and_caveats}

\begin{figure}
\centering
\includegraphics[width=0.49\textwidth]{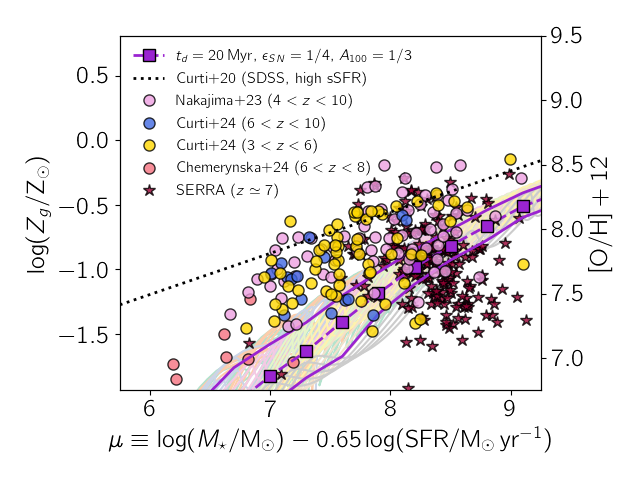}
\caption{
Deviation from the fundamental mass-metallicity relation (FMR) for high-$z$ galaxies.
We adopt the $\mu \equiv \log(M_\star/\msun) - 0.65 \log({\rm SFR}/\msunyr)$ FMR obtained by \citet{curti:2020} for the high sSFR galaxies sub-sample from the SDSS \citep[$z\lsim 2$][]{york:2000}.
We report the results of the weak feedback ($\epsilon_{\rm SN}=1/4$), modulated accretion ($\acosmo = 1/3$) models along with the data from JWST observations \citep{nakajima:2023,curti:2024,chemerynska:2024_b} and \code{SERRA} simulations \citep{pallottini:2022}.
\label{fig:FMR_flickering_impact}
}
\end{figure}

The minimal models presented here can also be useful as a tool to guide the development and compare complex cosmological simulations, particularly when focusing on relations that - like the MZR - are very sensitive to small variations of the underlying physics, that is included via sub-grid models in galaxy simulations \citep{agertz:2020}.

\subsection{Comparing numerical simulations}\label{sec:sub:simulations_comparison}

%
%
In galaxy simulations, ejective feedback, such as outflows caused by mechanical/kinetic prescriptions, can effectively suppress the SFR activity as much as preventative feedback, such as turbulence/delayed cooling prescriptions \citep{rosdahl:2017}, leading to similar SFR histories when the same initial conditions are considered \citep{lupi:2020}.
Despite similar SFR histories, other properties can differ significantly \citep{rosdahl:2017}, such as the dynamical state of a galaxy.

In the minimal model, $\epsilon_{SN}$ regulates the level of SFR suppression and determines the MZR normalization. In simulations, the level of SFR suppression can be quantified by the $M_\star-M_h$ relation.
\code{fire-2} predicts a $M_\star-M_h$ similar to \citet{behroozi:2013_c}; \code{SERRA} is also consistent with \citet{behroozi:2013_c} for $M_h\gsim 10^{11}\msun$ ($M_\star\gsim 10^9 \msun$), but on average\footnote{When $M_\star\lsim 10^6 \msun$, \code{SERRA} galaxies can be temporarily below the $M_\star-M_h$ relation from \citet{behroozi:2013_c} because of high radiation field and low dust content, which cause an efficient H$_2$ photodissociation, see the Alyssum effect in \citet{pallottini:2022}.} tends to be above such relation for $M_h\lsim 10^{9}\msun$.
Indeed, \code{SERRA} and \code{fire-2} give the same normalization for the MZR (see Fig. \ref{fig:MZR_obs_and_serra}) in the stellar mass range where the $M_\star-M_h$ is similar, while for $10^7 \lsim M_\star/\msun \lsim 10^9$ \code{SERRA} predicts a higher MZR normalization.

In the minimal model, the SFR flickering is regulated by $\tdelay$, which determines the MZR scattering. In simulations, the SFR flickering is regulated by the adopted feedback type (ejective vs preventative).
The MZR scatter is similar in \code{SERRA} and \code{fire-2} \citep{marszewski:2024} only for $M_\star\gsim 10^9 \msun$ (see Fig. \ref{fig:delta_vs_mstar}).
This is not surprising, as a higher SFR flickering is reported in \code{fire-2} \citep{sun:2023} with respect to \code{serra} \citep{pallottini:2023}, which can have further consequences on top of UV variations.
Higher SFR flickering in \code{fire-2} is due to violent outflows \citep{hopkins:2018}, while the suppression of SFR due to the photo-dissociation of molecular hydrogen in \code{serra} \citep{pallottini:2019} can more gently modulate the SFR activity without drastically modifying the galaxy dynamics.
Indeed, \citet{fujimoto:2024} show that \code{FIRE-2} galaxies \citep{wetzel:2023} have a much lower incidence of disks compared to \code{serra} galaxies \citep{kohandel:2024} at $M_\star \gsim 5\times 10^8 \msun$, i.e. although the SFR regulation in the two simulations is similar.
Despite the simplicity of the minimal model presented, hints of such differences in the simulations can be highlighted and explained.

%
%
However, it is difficult to directly quantify the differences in the simulations with the presented models.
%
%
For instance, the loading factor in our simplified models (but also in relatively more complex ones as \citealt{thompson:2016} and \citealt{pizzati:2020}) is a scalar quantity used as an input value for each galaxy, while in simulations is usually defined as an output that has a spatial dependence \citep{gallerani:2018}. 
Additionally, a precise quantification can be difficult; for instance, the MZR from \code{fire-2} \citep{marszewski:2024} is about 0.3 dex higher than the MZR from \code{fire} \citep{ma:2016}; in both cases, the slope is steeper (normalization is lower) than what observed by \citet{nakajima:2023,curti:2024} and is instead more consistent with \citet{chemerynska:2024_b}, however, the two simulations have the same loading factor \citep{pandya:2021}, thus it is difficult to explain the difference only in terms of our minimal models.
Moreover, when dealing with simulation snapshots, quantifying both the SFR event and the consequent outflows is tricky because of the time delay between SFR and feedback \citep{pandya:2021}, particularly for galaxies far from a quasi-equilibrium situation.

%
%
Further, the simplifications done in our minimal model impose we can only capture part of the complexity present in simulations.
For instance, while the SN-feedback delay is uniquely defined, simulations find a distribution of $\tdelay$ that depends on the underlying physical models included \citep{pallottini:2023}; thus, trying to find the set of parameters of a minimal model that gives the best match to a simulation will only give \textit{effective} parameters, a sort of summary statistic that can be used to compare different simulations.

Apart from these difficulties, a systematic parameter fitting procedure should give a better quantification of the comparison between different simulations, and can also be used to enable an even better match between the data and the minimal models. However, such models lack a few ingredients that prevent reproducing the full connection between metal build-up and the star formation history of galaxies.

\subsection{Caveats of the minimal model}\label{sec:sub:caveats}

The limitations of the presented minimal models can be appreciated by considering the fundamental mass-metallicity relation (FMR), i.e. the link between $Z_g$, $M_\star$, and SFR that is observed to be redshift independent up to $z\sim3$ \citep{mannucci:2010, curti:2020} and seems to break down at $z\gsim5$ \citep{nakajima:2023,curti:2024}.
We adopt the parameterization from \citet{curti:2020}
\begin{equation}
\mu \equiv \log(M_\star/\msun) - 0.65 \log({\rm SFR}/\msunyr)\,,
\end{equation}
which is obtained by minimizing the $Z_g$ dispersion for the high sSFR sub-sample from the Sloan Digital Sky Survey (SDSS, \citealt{york:2000}, $z\lsim 2$).

In Fig. \ref{fig:FMR_flickering_impact} we show the FMR for the minimal models matching the mean MZR and its dispersion, i.e. the weak feedback ($\epsilon_{\rm SN}=1/4$) set with modulated accretion ($\acosmo= 1/3$).
These models are below the SDSS relation by about 0.5 dex at $\mu \lsim 7$, with the separation that decreases with increasing $\mu$, down to 0.1 dex at $\mu\gsim 9$, almost connecting with the FMR trend observed at low $z$.
As expected \citep{nakajima:2023,curti:2024}, JWST galaxies are offset below the SDSS relation.
The best matching models for the MZR only partially match the JWST data in the FMR plane; the minimal models are closer to the JWST data when considering the lensed galaxies from \citet{chemerynska:2024_b}.

Further, the minimal models only produce downward deviation with respect to the FMR from the SDSS.
Instead, \code{serra} galaxies manage to better recover the trend for JWST galaxies around $\mu\simeq 8$, reproducing both downward and upward deviations with respect to the FMR from the SDSS; however, \code{serra} galaxies show an overabundance of low $Z_g$ galaxies at $\mu \simeq 8.5-9.0$, which is not seen in the JWST data.

In an equilibrium model \citep{lilly:2013}, the shape of the MZR is independent from the SFR, which uniquely determines the speed a galaxy can climb up the $Z_g$-$M_\star$ curve.
Note that the simplified model seems to deviate more from the data at low $\mu$, i.e. a combination of relatively high $\rm SFR$ and low $M_\star$; the adopted feedback increases in efficiency for lower $M_\star$ (eq. \ref{eq:outflow}).
Thus, the observed FMR deviation might be recovered by the minimal model if we consider not only ejecting but also preventing feedback, e.g. photodissociation of molecular hydrogen, which can suppress the star formation without affecting the gas mass (e.g. see Alyssum, a \code{serra} galaxy described in \citealt{pallottini:2022}) and can play a role in explaining some of the bright blue monsters seen by JWST \citep{ferrara:2024}.

As an alternative to keep matching the MZR and recovering the offset from the FMR, we could try to change the simplified SFR prescription (eq. \ref{eq:sfr}); this can be done for instance by adopting as the time scale for star formation a redshift dependent depletion time \citep{tacconi:2010,tacconi:2020,sommovigo:2022}, which are typically shorter than the selected $t_\star=1\,\rm Gyr$ by about a factor $\simeq 10$, similar to what \citet[][]{vallini:2024} report for $z\simeq 7$ galaxies by using \code{GLAM} \citep{vallini:2020, vallini:2021} determinations, thus promoting a faster galaxy evolution at high-$z$.

Indeed, by adopting the depletion time from \citet{sommovigo:2022} as the time scale for SFR (see App. \ref{app:sec:modify_sfr}, in particular, Fig. \ref{fig:analytical_model_examples_modified_tdepletion}), the stellar mass build-up is faster at $M_\star \lsim 10^{6} \msun$, while later saturates to values similar to those resulting from the $t_\star=1\,\rm Gyr$ case as both SFHs become feedback regulated, approaching a bathtub equilibrium.

However, on the one hand, the modifications affect the predictions of the minimal model mostly in the $M_\star \lsim 10^{6} \msun$ range; on the other hand, a $z$ average depletion time is sensitive to the population of the sample, e.g. for main sequence is $\simeq \times 5$ lower than for starburst \citep{tacconi:2020}, and the results from \citet{sommovigo:2022} are mostly relying on the data of $M_\star \gsim 10^{8-10} \msun$ galaxies from the ALPINE \citep{le_fevre:2019} and REBELS \citep{bouwens:2022} surveys.
To summarize, while trying to reconcile the FMR behavior is an interesting perspective and there are a few options, some care should be taken in extending/modifying the prescriptions for the minimal models; we leave such a possibility for future work.

\section{Conclusions}\label{sec:conclusions}

%
%
The exquisite spectroscopic data collected by JWST allow for the first time to derive the mass-metallicity relation (MZR) in galaxies up to redshift $z\simeq 10$.

%
%
In this work, we considered the MZR data for a combined sample of about 180 galaxies with stellar mass $10^6 \msun \lsim M_\star \lsim 10^{10} \msun$ at $z=3-10$ given \citet{nakajima:2023}, \citet{curti:2024}, and \citet{chemerynska:2024_b}, and obtained from the CEERS \citep{finkelstein:2023}, JADES \citep{bunker:2023}, and UNCOVER \citep{bezanson:2022} surveys, respectively. 

We compared the observations with the predictions from the \code{serra} simulations \citep{pallottini:2022}, finding a broad agreement with the observed $Z_g-M_\star$ data. However, simulations show a lack of a clear metallicity trend (flat MZR), and some tension, particularly at $M_\star\simeq 10^{10}\msun$.
%
%

To better understand the observed high-$z$ MZR and to clarify the behavior of the simulations, we have devised a minimal physical model for galaxy evolution in which star formation is fueled by cosmic gas accretion \citep{correa:2015_b} and is regulated by SN-driven outflows (eq. \ref{eq:outflow}), with a loading factor from \citet{muratov:2015} that is controlled by an efficiency $\epsilon_{\rm SN}$; additionally, we incorporated an explicit delay ($\tdelay$) between star formation and supernova feedback, inducing a stochastic SFR behavior. Some models also explore the possible modulation of cosmic accretion (eq. \ref{eq:mhalo_growth_modified}). 
%
%
The main results are:
\begin{itemize}
  \item[$\bullet$] To recover the average trend of the MZR observed at high-$z$, outflows must be less efficient ($\epsilon_{\rm SN} = 1/4$) than predicted in \citet{muratov:2015}. However, this evidence is consistent with the mass loading factors inferred for JADES galaxies \citep{carniani:2024}.
  \item[$\bullet$] To match the MZR dispersion in the full stellar mass range, both a delayed ($\tdelay=20\,\myr$) SN feedback, and modulation of the cosmic accretion ($\acosmo = 1/3$) are required.
  \item[$\bullet$] The r.m.s. variation of the MZR ($\sigmaz$) is very sensitive to the SFR flickering ($\sigmasfr$). The weak feedback necessary to reproduce the average MZR trend ($\epsilon_{\rm SN} = 1/4$) also results in a low-amplitude SFR flickering ($\sigmasfr\simeq 0.2$). This prescription correctly reproduces the moderate MZR scatter ($\sigmaz\simeq 0.25$) observed by JWST.
  \item[$\bullet$] \code{serra} galaxies have slightly higher $\sigmasfr\simeq 0.24$ \citep{pallottini:2023}, resulting in a metallicity scatter of $\sigmaz\simeq 0.3$, higher than observed. Differently from the analytical model, \code{serra} galaxies also overpredict the number of low metallicity ($Z_g\simeq 0.2\, \zsun$) galaxies at $M_\star\simeq 10^{10}\msun$.
  \item[$\bullet$] In general, any model predicting $\sigmasfr\gsim 0.5$ is likely to overshoot the observed metallicity r.m.s. scatter, possibly leading to a \quotes{chemical chaos} ($\sigmaz \gsim 1.4$) that is not present in JWST data, for which $\sigmaz\simeq 0.25$.
\end{itemize}

%
%
The last point also entails that simultaneously explaining the observed MZR \citep{nakajima:2023,curti:2024} and the overabundance of bright galaxies observed by JWST \citep[][]{finkelstein:2022,naidu:2022} through high SFR flickering \citep{mason:2023,mirocha:2023,shen:2023,sun:2023_b,kravtsov:2024} is very challenging, considering that a $\sigmasfr\simeq 0.8$ is required to explain $z>10$ observations \citep[][]{munoz:2023}.
We note that while both the SN-feedback delay and cosmic accretion modulation induce SFR flickering, the latter tends to give a gentler modulation of $Z_g$, since high feedback efficiency and a delay are needed in order to off-balance the MZR generated from a bathtub-like equilibrium \citep{lilly:2013}.
However, variation of cosmic accretion can only go up to 0.3 dex \citep{rodriguez_puebla:2016, ren:2019, mirocha:2021}. Thus, alternative explanations \citep{dekel:2023,ferrara:2023} seem more favored to solve the overabundance problem.

As noted in the Introduction\footref{footnote:face_value}, we caution that we are taking the observed MZR scattering at face value. However, this quantity might be affected in a non-trivial way by uncertainties in the $Z_g$ determinations.
However, observationally the MZR has been explored by different groups, samples, and methods, i.e. metallicity determinations are available with both auroral lines that can be used to directly compute the electron temperature \citep{morishita:2024}, and indirect methods based on combinations of various calibrators \citep{heintz:2023,nakajima:2023,curti:2024}. It is reassuring to see that these works and datasets independently report a similar scattering of $\simeq 0.3-0.5$ dex.
For this reason, we consider the current measurements of the MZR dispersion as relatively robust. In the future, it is likely that the dispersion might decrease, given that the current sample size is relatively small ($\sim 200$ galaxies) and covers a relatively large redshift range ($z\simeq 3-10$).

%
%
While we showed that the presented minimal model can also be used to guide our intuition in comparing complex cosmological simulations, we recall that such a simplified model does not fully capture the physical complexity of the connection between galaxy formation and metal enrichment, as highlighted by the poor match with the JWST data when analyzing the off-set from the fundamental mass-metallicity relation.
%
%
Combining insights from such models with further analysis of ongoing observations, and more complex numerical simulations is crucial for understanding the galaxy formation and evolution process at high-$z$.

\begin{acknowledgements}
We acknowledge the CINECA award under the ISCRA initiative, for the availability of high-performance computing resources and support from the Class B project SERRA HP10BPUZ8F (PI: Pallottini).
AF acknowledges support from the ERC Advanced Grant INTERSTELLAR H2020/740120 (PI: Ferrara).
SC and GV acknowledge support support from the ERC Starting Grant WINGS H2020/101040227 (PI: Carniani).
Any dissemination of results must indicate that it reflects only the author's view and that the Commission is not responsible for any use that may be made of the information it contains.
Partial support (AF) from the Carl Friedrich von Siemens-Forschungspreis der Alexander von Humboldt-Stiftung Research Award is kindly acknowledged.
We gratefully acknowledge the computational resources of the Center for High Performance Computing (CHPC) at SNS.
We acknowledge usage of the Python programming language \citep{python2,python3}, Astropy \citep{astropy}, Cython \citep{cython}, Matplotlib \citep{matplotlib}, Numba \citep{numba}, NumPy \citep{numpy}, \code{pynbody} \citep{pynbody}, and SciPy \citep{scipy}.
\end{acknowledgements}

\bibliographystyle{aa_url}
\bibliography{bib/master, bib/codes}

\appendix


\section{Modifications to the SFR time scales}\label{app:sec:modify_sfr}

\begin{figure*}
\centering
\includegraphics[width=0.49\textwidth]{plots/mass_comparison.png}
\includegraphics[width=0.49\textwidth]{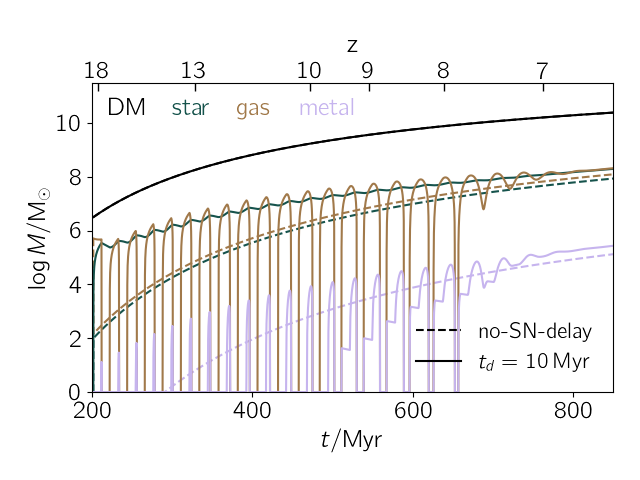}
\caption{
Difference in the evolution of a $M_{dm} = 3\times 10^{10}\msun$ computed by prescribing the fiducial ($t_\star = 1\, \rm Gry$, \textbf{left panel}) and faster ($t_\star^{mod}$, eq. \ref{eq:t_dep_sommovigo}, \textbf{right panel}) star formation timescale (eq. \ref{eq:sfr}).
For both panels, Dark matter, stellar, gas, and metal mass evolution is reported for both the no-SN-delay and $\tdelay=10\,\myr$ model with the same notation of as in Fig. \ref{fig:analytical_model_examples}.
Note that the left panel is exactly Fig. \ref{fig:analytical_model_examples}, it is reported here to have a clearer visual comparison.
\label{fig:analytical_model_examples_modified_tdepletion}
}
\end{figure*}

In bathtub models \citep{lilly:2013}, the shape of the MZR is mostly independent from the functional form of the SFR, as the equilibrium solution is determined by the balance between infall and outflow, while the SFR uniquely determines the speed a model climbs the MZR from low to high $M_\star$.
Instead, the SFR prescription can be important if a stochastic behavior is present in the system and when analyzing the FMR.

In the fiducial model presented in the paper we assumed that the SFR is proportional to the gas mass and is regulated by a fixed time scale for star formation, i.e. $t_\star = 1\,\rm Gyr$. As a natural alternative for a fixed $t_\star$, a model can consider adopting $t_\star$ via empirical determinations \citep{tacconi:2010} or model-based predictions \citep{sommovigo:2022} of the depletion time, which is expected to decrease with redshift, thus giving a faster star formation rate at higher $z$.

As a test, we modify our models by changing $t_\star$ in our SFR prescription (eq. \ref{eq:sfr}) by adopting the scaling suggested by \citet[][see eq. 16 therein]{sommovigo:2022}
\begin{align}
\label{eq:t_dep_sommovigo}
t_\star^{mod} &\simeq \frac{8.95 \, \rm Gyr}{E(z)} \\
&= \frac{8.95 \, \rm Gyr}{ [-0.24 + 0.75 (1+z)]\sqrt{\Omega_m(1+z)^3 + \Omega_\Lambda}}\,,\nonumber
\end{align}
and setting the proportionality constant in order to match \citet{tacconi:2010} determinations at $z=0$.
As a reference, $t_\star \simeq 20\, t_\star^{mod}$ at $z\simeq 10$, thus a faster stellar mass build is expected for the modified model. 

As a benchmark, we select the halo with $M_{dm}=3\times 10^{10}\msun$ at $z=6$ analyzed in Sec. \ref{sec:model_overview} (see Fig. \ref{fig:analytical_model_examples}), evolve it with $t_\star = 1\,\rm Gyr$ and $t_\star^{mod}$ from (eq. \ref{eq:t_dep_sommovigo}), and plot the results in Fig. \ref{fig:analytical_model_examples_modified_tdepletion} for both $\tdelay=10\,\myr$ and a reference no-SN-delay case.

The largest differences between the $t_\star = 1\,\rm Gyr$ and $t_\star^{mod}$ models are present for the stellar mass build up at $z\gsim 10$, when the galaxy has $M_\star \lsim 10^7\msun$,
The $t_\star^{mod}$ model promotes a faster build-up, particularly when a SN-delay is present, as the lack of feedback combined with the short SFR timescales makes the galaxy reach $M_\star \lsim 10^6\msun$ during the first burst.
As time progresses, differences between the models tend to wash out, as the galaxy approaches $M_\star \simeq 10^8\msun$, the SFH starts to become feedback-regulated, and at $z\simeq 6$, stellar, gas, and metal masses are roughly similar in all cases. 
Interestingly, while the amplitude of the flickering is similar ($\sigmasfr\simeq 0.5$), the modulation acts on slightly shorter time scales for the $t_\star^{mod}$ model, as a consequence of the faster SFR rise and the heavier off-balance caused by the stellar feedback.
This implies that the $t_\star^{mod}$ model at $z\lsim 7.5$ gives an even enhanced dispersion for the MZR (from $\sigmaz\simeq 0.7$ to $\sigmaz\simeq 1.7$).
Note that when no-SN-delay is present, the SFH evolution is feedback regulated from the start, and the main difference is an increased gas fraction for the $t_\star = 1\,\rm Gyr$ model.

Note that adopting \citet{sommovigo:2022} implies a depletion time with a $\propto z^{-2.5}$ redshift dependence, i.e. steeper with respect to the $\propto z^{-1.5}$ dependence from \citet{tacconi:2010}. Thus, by assuming \citet{tacconi:2010} instead of \citet{sommovigo:2022} to modify the SFR timescale, we expect smaller variation with respect to the fiducial $t_\star=1\,\rm Gyr$ model.

\section{MZR: effect of time sampling}\label{app:sec:coraser_dt}

\begin{figure}
\centering
\includegraphics[width=0.49\textwidth]{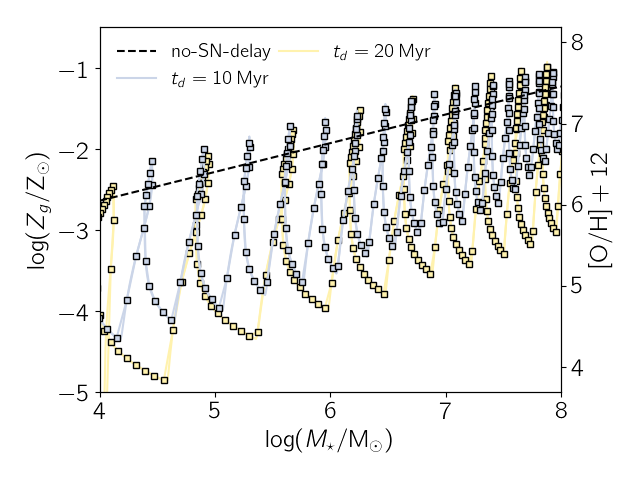}
\caption{
Impact of a coarse time sampling in determining the burstiness on the MZR for a system with $M_{dm} = 3\times 10^{10}\msun$ at $z=6$ evolved with our minimal physical model (cfr with Fig. \ref{fig:analytical_model_MZR_examples}).
Different lines indicate different delay times for the SN feedback, as indicated in the legend.
Squares (continuous lines) indicate a $2\,\myr$ ($0.2\,\myr$) time sampling.
For visualization sake, the number of models is less than that present in Fig. \ref{fig:analytical_model_MZR_examples}.
\label{fig:app:analytical_model_MZR_examples_coarse}
}
\end{figure}

While dedicated work is needed to properly forward model an observational-like metallicity and the associated scattering, here we touch upon the effect of time sampling.
The scatter that is actually measurable depends on the timescale that a galaxy is both observable and persists in a particular state.

To have an idea of such an effect, in Fig. \ref{fig:app:analytical_model_MZR_examples_coarse} we report the evolution of the MZR from our target $M_{dm} = 3\times 10^{10}\msun$ system at $z=6$ for different $\tdelay$ (see full discussion in Sec. \ref{sec:model_overview}, particularly referring to Fig. \ref{fig:analytical_model_MZR_examples}) by comparing the idealized $0.2\,\myr$ time sampling with a more realistic $2\,\myr$ time sampling.
Considering a single MZR loop induced by a star formation burst, the $2\,\myr$ time sampling better catches the decrease of $Z_g$ rather than the increase of $Z_g$. The coarse times sampling more closely recovers the ideal trend at higher stellar masses. The derived scattering is mostly unaffected.

\end{document}